\documentstyle[epsfig]{mn}

\newif\ifAMStwofonts
\ifoldfss
  \ifCUPmtlplainloaded \else
    \NewTextAlphabet{textbfit} {cmbxti10} {}
    \NewTextAlphabet{textbfss} {cmssbx10} {}
    \NewMathAlphabet{mathbfit} {cmbxti10} {} % for math mode
    \NewMathAlphabet{mathbfss} {cmssbx10} {} %  "   "    "
  \fi
  \ifAMStwofonts
    \ifCUPmtlplainloaded \else
      \NewSymbolFont{upmath} {eurm10}
      \NewSymbolFont{AMSa} {msam10}
      \NewMathSymbol{\upi}     {0}{upmath}{19}
      \NewMathSymbol{\umu}     {0}{upmath}{16}
      \NewMathSymbol{\upartial}{0}{upmath}{40}
      \NewMathSymbol{\leqslant}{3}{AMSa}{36}
      \NewMathSymbol{\geqslant}{3}{AMSa}{3E}

    \fi
  \fi
\fi % End of OFSS

\ifnfssone
  \newmathalphabet{\mathit}
  \addtoversion{normal}{\mathit}{cmr}{m}{it}
  \addtoversion{bold}{\mathit}{cmr}{bx}{it}
  \newmathalphabet{\mathbfit} % math mode version of \textbfit{..}
  \addtoversion{normal}{\mathbfit}{cmr}{bx}{it}
  \addtoversion{bold}{\mathbfit}{cmr}{bx}{it}
  \newmathalphabet{\mathbfss} % math mode version of \textbfss{..}
  \addtoversion{normal}{\mathbfss}{cmss}{bx}{n}
  \addtoversion{bold}{\mathbfss}{cmss}{bx}{n}
  \ifAMStwofonts
    \ifCUPmtlplainloaded \else
      \UseAMStwoboldmath
      \makeatletter
      \new@mathgroup\upmath@group
      \define@mathgroup\mv@normal\upmath@group{eur}{m}{n}
      \define@mathgroup\mv@bold\upmath@group{eur}{b}{n}
      \edef\UPM{\hexnumber\upmath@group}
      \new@mathgroup\amsa@group
      \define@mathgroup\mv@normal\amsa@group{msa}{m}{n}
      \define@mathgroup\mv@bold\amsa@group{msa}{m}{n}
      \edef\AMSa{\hexnumber\amsa@group}
      \makeatother
      \mathchardef\upi="0\UPM19
      \mathchardef\umu="0\UPM16
      \mathchardef\upartial="0\UPM40
      \mathchardef\leqslant="3\AMSa36
      \mathchardef\geqslant="3\AMSa3E
    \fi
  \fi
\fi % End of NFSS release 1

\ifnfsstwo
  \DeclareMathAlphabet{\mathbfit}{OT1}{cmr}{bx}{it}
  \SetMathAlphabet\mathbfit{bold}{OT1}{cmr}{bx}{it}
  \DeclareMathAlphabet{\mathbfss}{OT1}{cmss}{bx}{n}
  \SetMathAlphabet\mathbfss{bold}{OT1}{cmss}{bx}{n}
  \ifAMStwofonts
    \ifCUPmtlplainloaded \else
      \DeclareSymbolFont{UPM}{U}{eur}{m}{n}
      \SetSymbolFont{UPM}{bold}{U}{eur}{b}{n}
      \DeclareSymbolFont{AMSa}{U}{msa}{m}{n}
      \DeclareMathSymbol{\upi}{0}{UPM}{"19}
      \DeclareMathSymbol{\umu}{0}{UPM}{"16}
      \DeclareMathSymbol{\upartial}{0}{UPM}{"40}
      \DeclareMathSymbol{\leqslant}{3}{AMSa}{"36}
      \DeclareMathSymbol{\geqslant}{3}{AMSa}{"3E}
    \fi
  \fi
\fi % End of NFSS release 2

\ifCUPmtlplainloaded \else
  \ifAMStwofonts \else % If no AMS fonts
    \def\upi{\pi}
    \def\umu{\mu}
    \def\upartial{\partial}
  \fi
\fi

\newcommand{\Msol}{M$_\odot$}
\title{Prospects for Type Ia Supernova explosion mechanism identification
with $\gamma$-rays}

\author[J. G\'{o}mez-Gomar, J. Isern and P. Jean ]
       {Jordi G\'{o}mez-Gomar $^1$, Jordi Isern $^1$\and Pierre Jean$^2$ \\
$^1$Institut d'Estudis Espacials de Catalunya (IEEC), CSIC, Barcelona,
 Gran Capit\`a 2 -- 4, C.P. 08034, Spain.\\
$^2$CESR (CNRS), av. du Colonel Roche, BP 4346, 31028 Toulouse Cedex, France.}

\date{Accepted \today.
      Received \today;
      in original \today}

\pagerange{\pageref{firstpage}--\pageref{lastpage}}

\begin{document}

\maketitle

\label{firstpage}

\begin{abstract}
The explosion mechanism associated with thermonuclear
supernovae (SNIa) is still a matter
of debate.  There is a wide  agreement  that 
 high amounts of of radioactive nuclei
are produced during  these events and they  are expected to be strong \(\gamma 
 \)-ray emitters.   
 In the past, several authors have  investigated the use of 
this $\gamma$-ray emission as a diagnostic tool.
 In this  paper we have done a complete
 study of the $\gamma$-ray spectra associated with all the 
different scenarios  currently proposed. This includes
 detonation, delayed detonation, deflagration and the off-center
 detonation. We have  performed
accurate simulations for this complete 
set of models in order to  determine the most promising spectral features
 that could
 be used to discriminate among the different models.
 Our study is not limited   to  qualitative arguments. Instead, we have
  quantified the differences among the spectra and
 established distance limits for their detection. The calculations have
 been performed considering the best current response estimations 
of  the SPI and IBIS instruments aboard INTEGRAL in such a way that  our
 results can  be used
  as a guideline to evaluate the capabilities of INTEGRAL in the study of
 type Ia supernovae. For the purpose of completeness we have also investigated 
the nuclear excitation and spallation reactions as a possible 
secondary source of
 $\gamma$-rays present in some supernova scenarios. We conclude that 
 this mechanism can be neglected due to its small 
contribution.

\end{abstract}


\section{Introduction}
It is commonly accepted that Type Ia  supernovae are the result of the thermonuclear 
explosion of a
 mass accreting CO white dwarf. In the outburst,  high amounts of
 $^{56}$Ni and other radioactive isotopes are produced  opening
 the possibility to use $\gamma$-rays as a diagnostic tool.   
  Although there is an  agreement about the basic properties of such
  supernovae, there are  several theories to account for these
 events. One class of  models assumes that the parent white dwarf is very
 close to the Chandrasekhar mass and that the thermonuclear runaway starts at the
 center. According to the properties of the burning front, three cases can be considered:
 detonation model \cite{Ar69}, deflagration
 model \cite{No84} and delayed detonation model 
\cite{Kh91}.  Recently, a fourth model based on
 a sub-Chandrasekhar mass progenitor has  gained  popularity due to
 some observational evidences (Phillips 1993; Maza et al. 1994; Hamuy et al. 1995).
In this case,
   the explosion is triggered by  the ignition of a freshly accreted He mantle
(Ruiz-Lapuente et al. 1993; Livne et al. 1993; Wossley et al. 1994; Arnett 1994).

Regarding the amount of radioactive material synthesized, the chemical  
composition and the density and velocity profiles, the
 properties of   ejecta are different  for each one of these models.
  These   differences affect the evolution of the total intensity of the
 $\gamma$-ray lines, their relative ratios and even their widths and shapes as well as the
 importance and extension of the continuum component of the spectrum.

Several authors have already investigated   the $\gamma$-ray
 emission of type Ia supernovae 
(Gehrels et al. 1987; Ambwani \& Burrows 1988; Burrows \& The 1990, Burrows 1991; The et al. 1993;
 Ruiz-Lapuente et al. 1993b; H\"oflich \& Khoklov 1994; Kumagai \& Nomoto 1995; 
Woosley \& Timmes 1996) for different SNIa models.  In most of these works a
 reduced number of SNeIa scenarios were studied, while in none of them a detailed
 determination of the detectability by INTEGRAL was performed. The aim of this
 paper is to accurately 
compute the evolution of the $\gamma$-ray spectra for a complete set  of
 models, covering all the theories already  mentioned, 
and to  determine which  spectral features could provide
 interesting  information about  SNIa.
 A Monte-Carlo $\gamma$-ray transfer
 code has been developed  to compute the $\gamma$-ray emission for all the
explosion models. Afterwards, the spectra have been convolved with the expected 
instrumental
 response  for IBIS \cite{Le95} and SPI \cite{Pj95} on-board of
 INTEGRAL  to obtain the observational properties.
A set of quantities that characterize the  detectable
 spectral properties including line and continuum intensities and line shapes 
have been determined.
 We have also computed which are the  detectability limits of these
 properties and investigated when, any given model  could be rejected or
 identified if a SNIa is observed.
 Although the radioactive decay of freshly synthesized nuclei
 is the main source of $\gamma$-ray emission for SNIa, it is not the unique one.
 We have also investigated the emission produced by the nuclear excitation due to the
  interaction between the fast ejecta and the circumstellar medium, as would happen in
 the case  of the explosion of a type Ia supernova in a symbiotic binary, or in a
 ISM cloud. Although, this mechanism is much weaker it  has the advantage 
that it  operates on longer time-scales (up to 1000's of years).

\section{Models and Results}
In order to compute the $\gamma$-ray spectra of the different models we have developed
 a code for the treatment of the $\gamma$-ray transfer, as described by
 Pozdnyakov et al. (1983) and Ambwani et al. (1988).
 It is based on the Monte-Carlo method technique, which allows the 
 treatment of  the comptonization process without approximations.
 With the code we can simulate the $\gamma$-ray spectra emitted by a
 SNIa  with arbitrary composition, velocity and density profiles.
 Although  many radioactive chains are included in the code only the following
 ones are important in the case of type Ia SNe:
$$^{56}Ni \rightarrow ^{56}Co \rightarrow ^{56}Fe$$
$$^{57}Ni \rightarrow ^{57}Co \rightarrow ^{57}Fe$$
Three different sources of opacity have been taken into account: Compton
 scattering, photo-electric absorption and e$^+$  e$^-$ pair production.
 The cross section for Compton scattering is given by the habitual 
Klein-Nishina expression, while absorption and pair production cross
 sections were taken from the compilation of experimentally evaluated data
 maintained by the Brookhaven National Laboratory

Three sources of $\gamma$ photons are considered besides nuclear decay: direct
emission of two photons (511 keV)   by electron positron annihilation, 
indirect emission of two or three photons by  positronium annihilation
\cite{Or49}, and emission of fluorescence 
photons (not relevant in the present work since they are low energy 
photons E $<$ 10 keV).
 Simulations have been carried out  to obtain both the evolution of the 
intensity   of the strongest lines in the form of light curves and 
 the detailed spectra at given times.

\begin{figure}
	\begin{center}
\epsfig{file=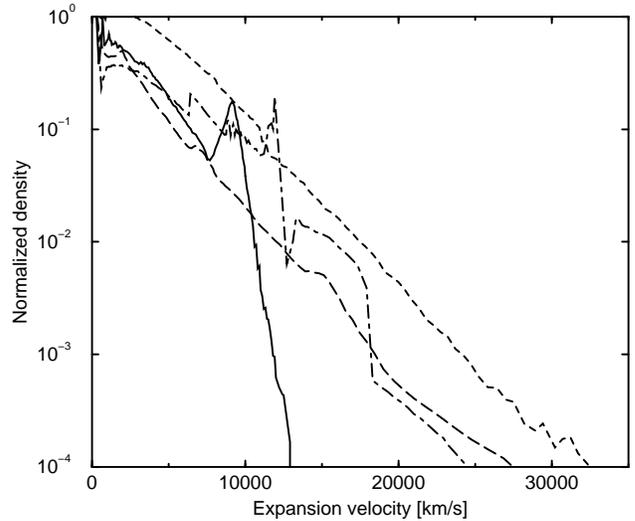,height=8 cm}
\end{center}
\caption{Density profiles as a function of the  velocity of the
 ejecta for different models during the homologous expansion phase. Solid
 line corresponds to DEF model, long-dashed line to DEL model, dashed line to
 DET model and dot-dashed line to SUB model.}
 \label{fig}
\end{figure}

 The  properties of the ejecta for the different models have been kindly 
provided by E. Bravo who obtained them from  accurate simulations of SNIa 
explosions. All the calculations were   performed following the evolution 
 of the system through the
accretion phase   and starting with a  0.8 \Msol 
partially cooled white dwarf with a composition  (X$_{C}$=0.51, X$_{O}$=0.49).
 The general procedure followed in the simulations is fully described in Bravo
 et al. (1996) with the exception of the sub-Chandrasekhar model 
(hereafter,SUB) were 
 a particularly accurate simulation of the accretion phase was carried out  by
Jos\'e (Jos\'e 1996) with a hydrodynamical code. 
This calculation is, up to date,
 the most consistent simulation of a sub-Chandrasekhar supernova in 1D.
 Three other  models have been considered:
 DEF, DEL and  DET, representing  deflagration, delayed detonation and
 detonation  supernovae, respectively.
  The details of the parameterizations adopted in the
 propagation of the burning front for  DEF, DEL, and DET models
 are also described in Bravo et al. 1996. The main properties  of these
models at the beginning of the homologous expansion phase are
summarized in Table \ref{Tab1} and Figure \ref{fig}. The values in the table 
correspond to  the $^{56}$Ni and $^{57}$Ni contain, mass of C + O, velocity of 
the shell with m= 1 \Msol and total kinetic energy. The basic properties of our
models (particularly, the ejected  $^{56}$Ni mass and kinetic energy) are compatible within
 uncertainities
 with those found  in the literature 
(see for example models cdtg7 \cite{Wo86} DF1 and DET1 \cite{Ho96},  N21
 \cite{Kh91}, Model 7 \cite{Wo94}).

\begin{table*}
\centering
\begin{minipage}{120 mm}
\caption
 {Main properties of the ejecta after the beginning of the homologous 
expansion phase.}
\label{Tab1}
\begin{tabular}{@{}lccccc}
Model &
$^{56}$Ni [\Msol] &
$^{57}$Ni [\Msol] &
C+O [\Msol]  &
v(1~ \Msol)[10$^9$ cm/s] &
E$_k$ [10$^{51}$ ergs] \\
DEF & 0.5 & 0.02 & 0.5 &0.9 & 0.7 \\
DEL & 0.8 & 0.02 & 0.02 & 1.1 & 1.7 \\
DET & 0.7 & 0.04 & $\sim$ 0 & 1.1 & 1.5 \\
SUB & 0.6 & 0.01 & 0.13 & 0.6 & 1   \\
\end{tabular}

\end{minipage}
\end{table*}

%% Results

\begin{figure}
	\begin{center}
\epsfig{file=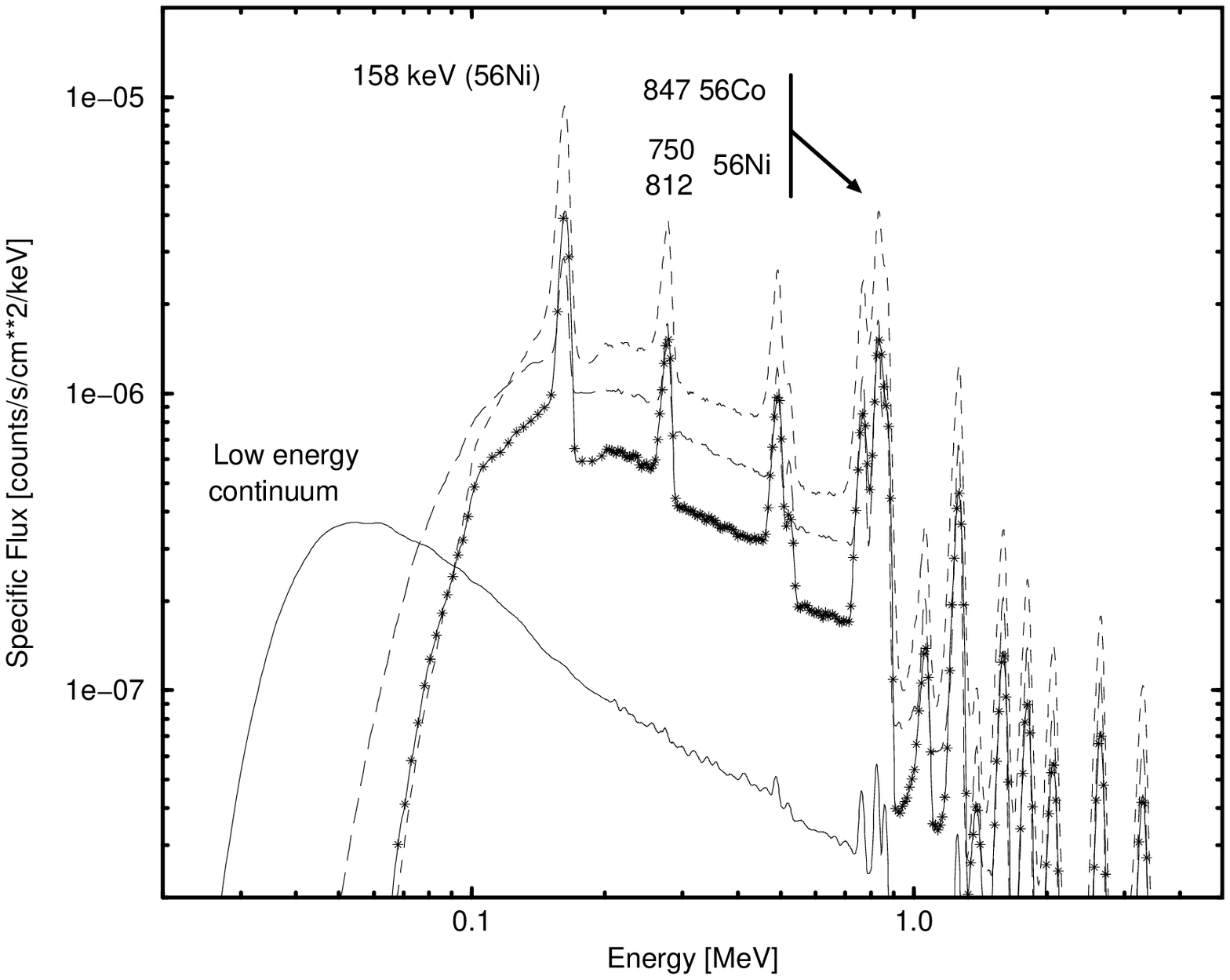,height=8 cm, width=8 cm}

\vspace{-1.5 cm}

\epsfig{file=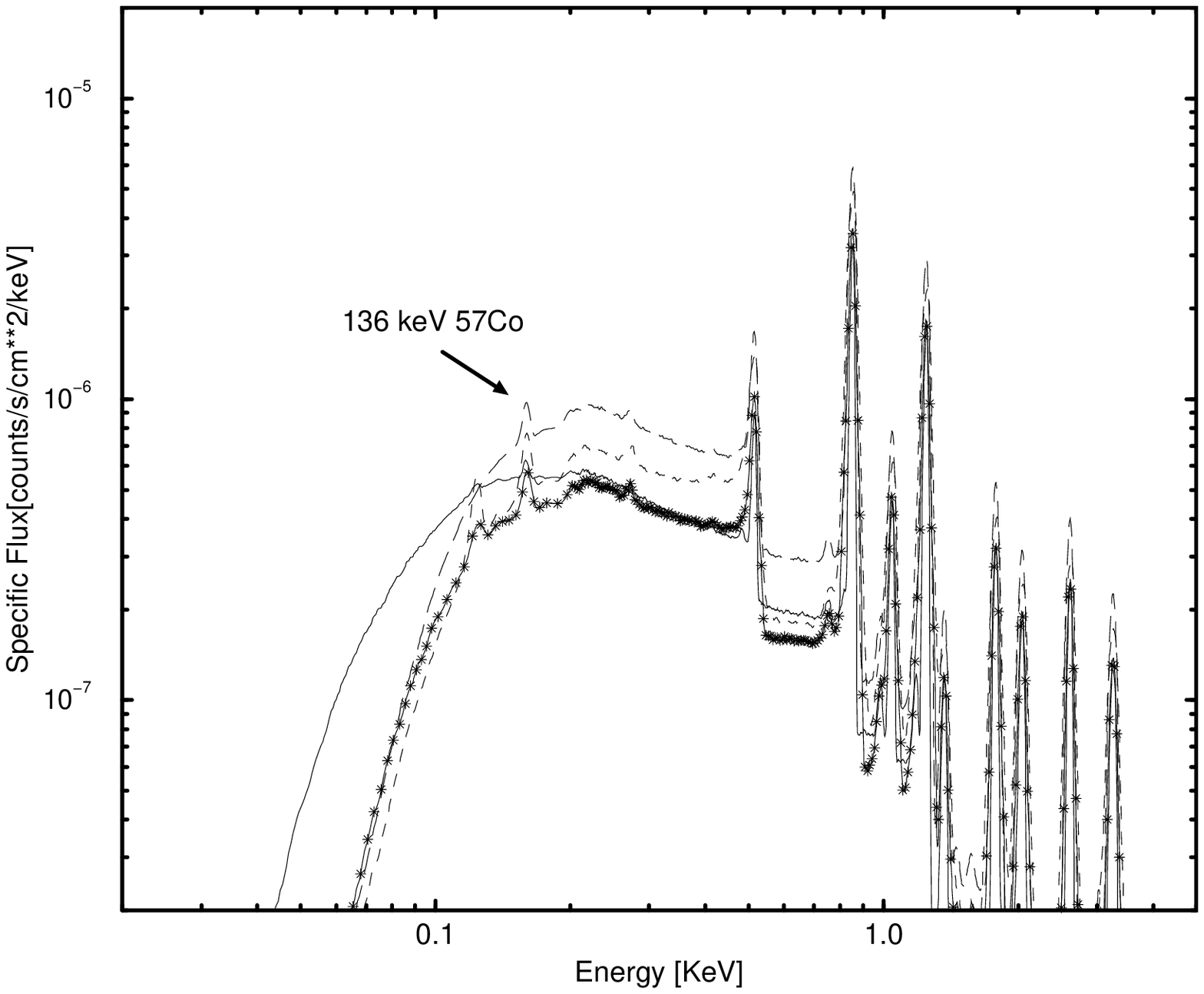,height=8 cm, width=8 cm}

\vspace{-1.5 cm}

\epsfig{file=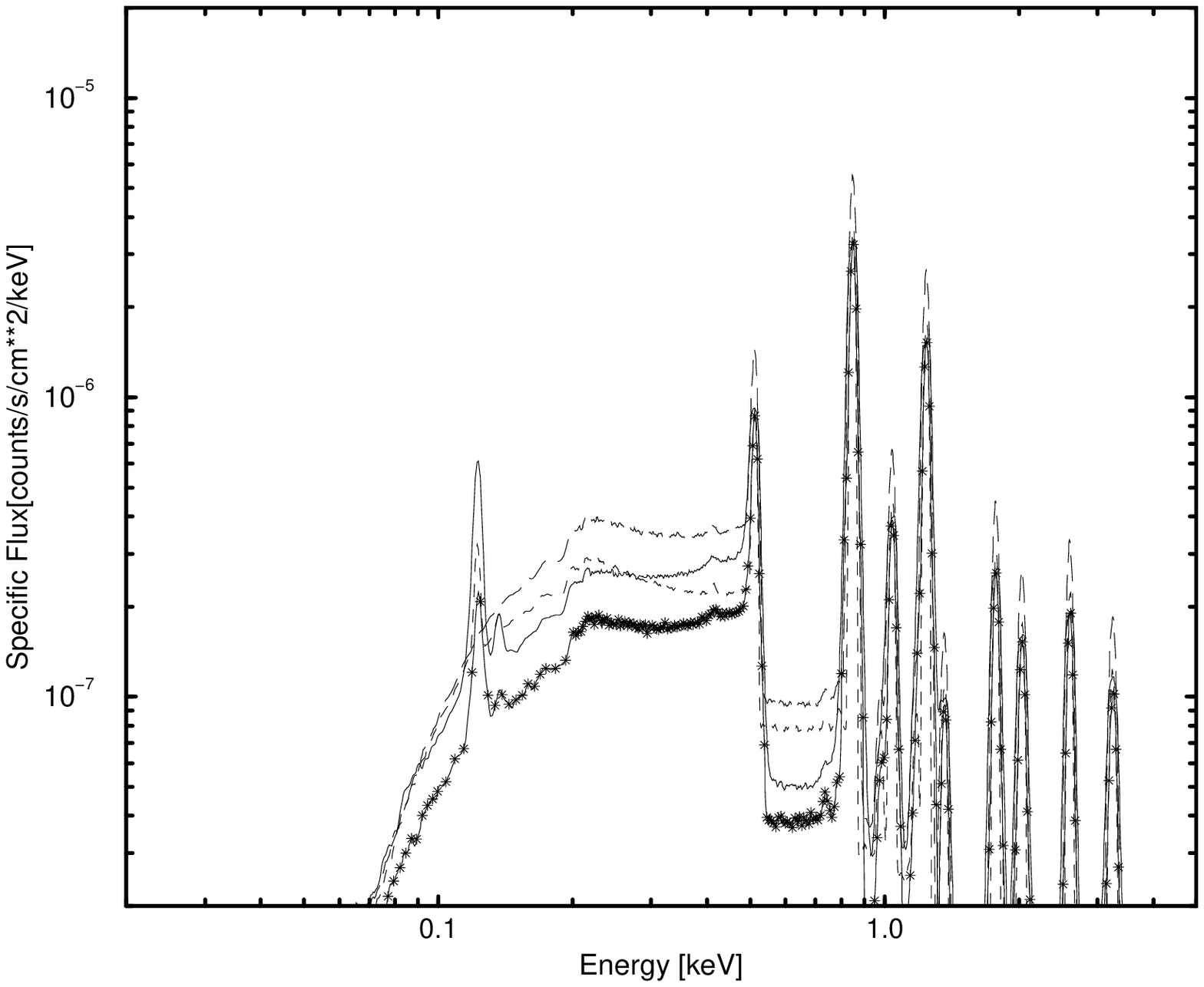,height=8 cm, width=8 cm}
	\end{center}
\caption{Spectral evolution for the four models  (5 Mpc)  at 
20 days (top panel), 60 days (middle panel) and 120 days (bottom panel)
 after the explosion. Solid
 lines correspond to DEF model, long-dashed lines to DEL model, dashed lines to
 DET model and starred-line to SUB model.}
\label{fig1}
\end{figure}

\begin{figure}
\begin{center}
\epsfig{file=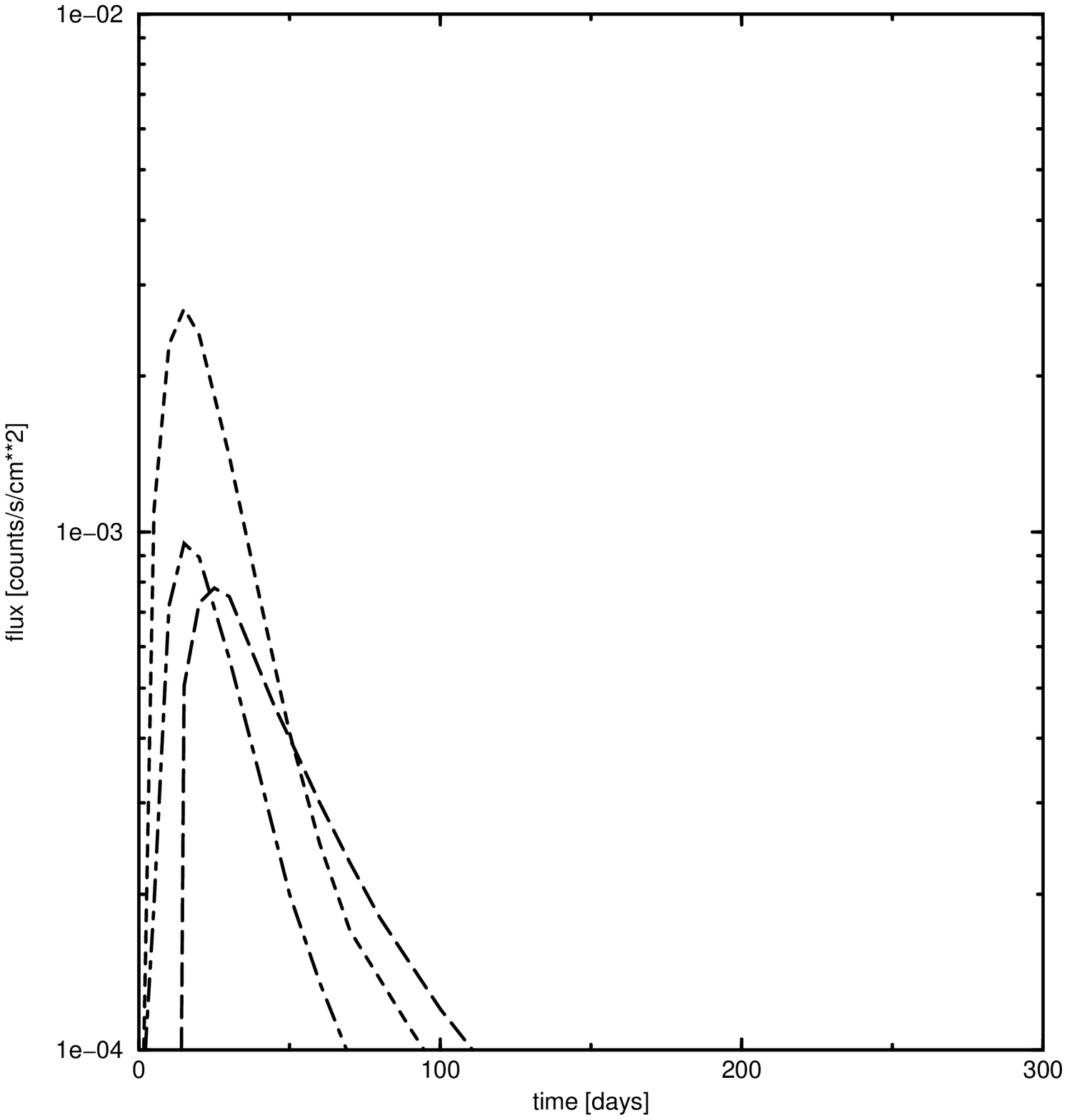,height=8 cm, width=8 cm}

\vspace{-1.5 cm}

\epsfig{file=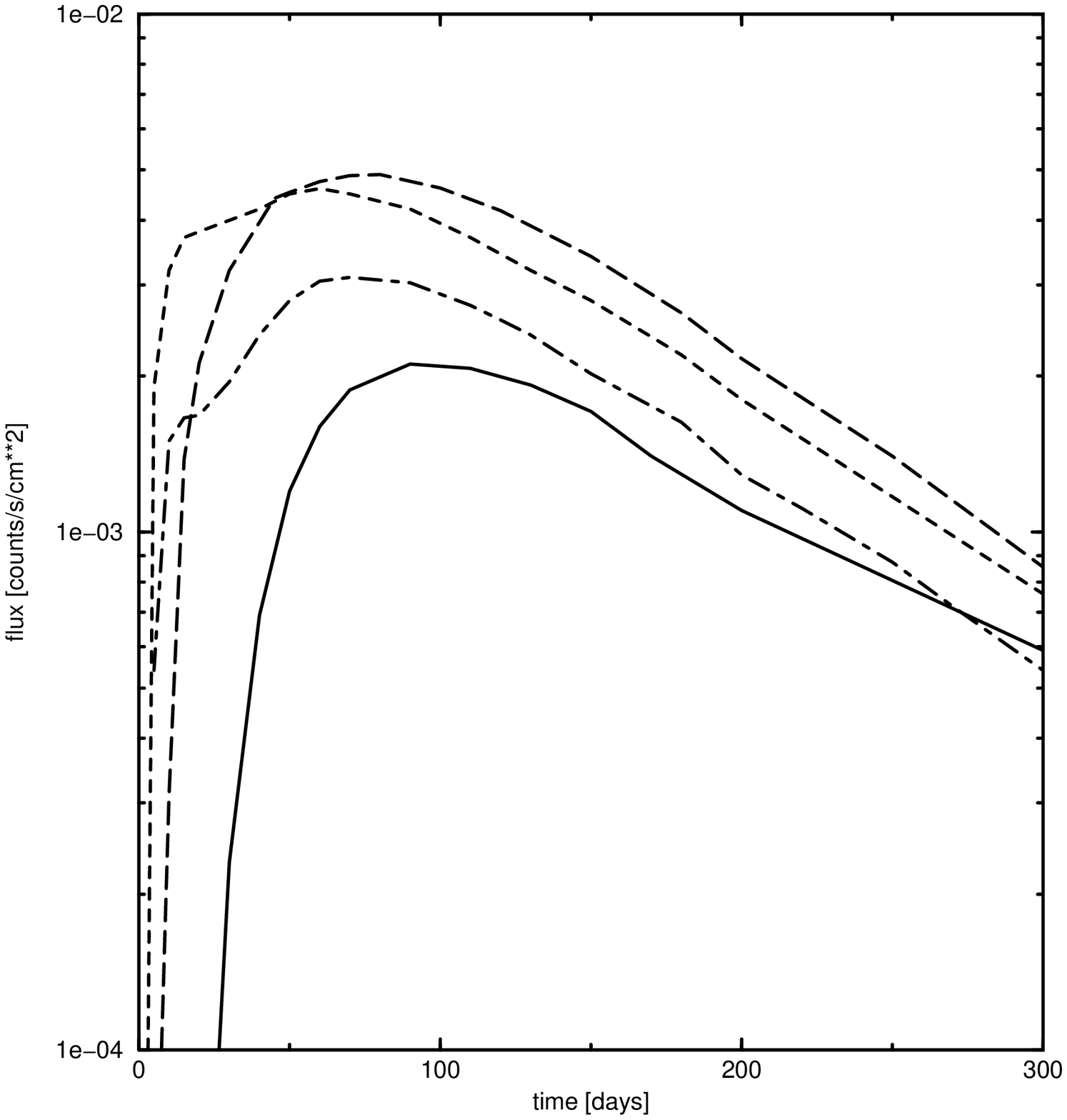,height=8 cm, width=8 cm}

\vspace{-1.5 cm}

\epsfig{file=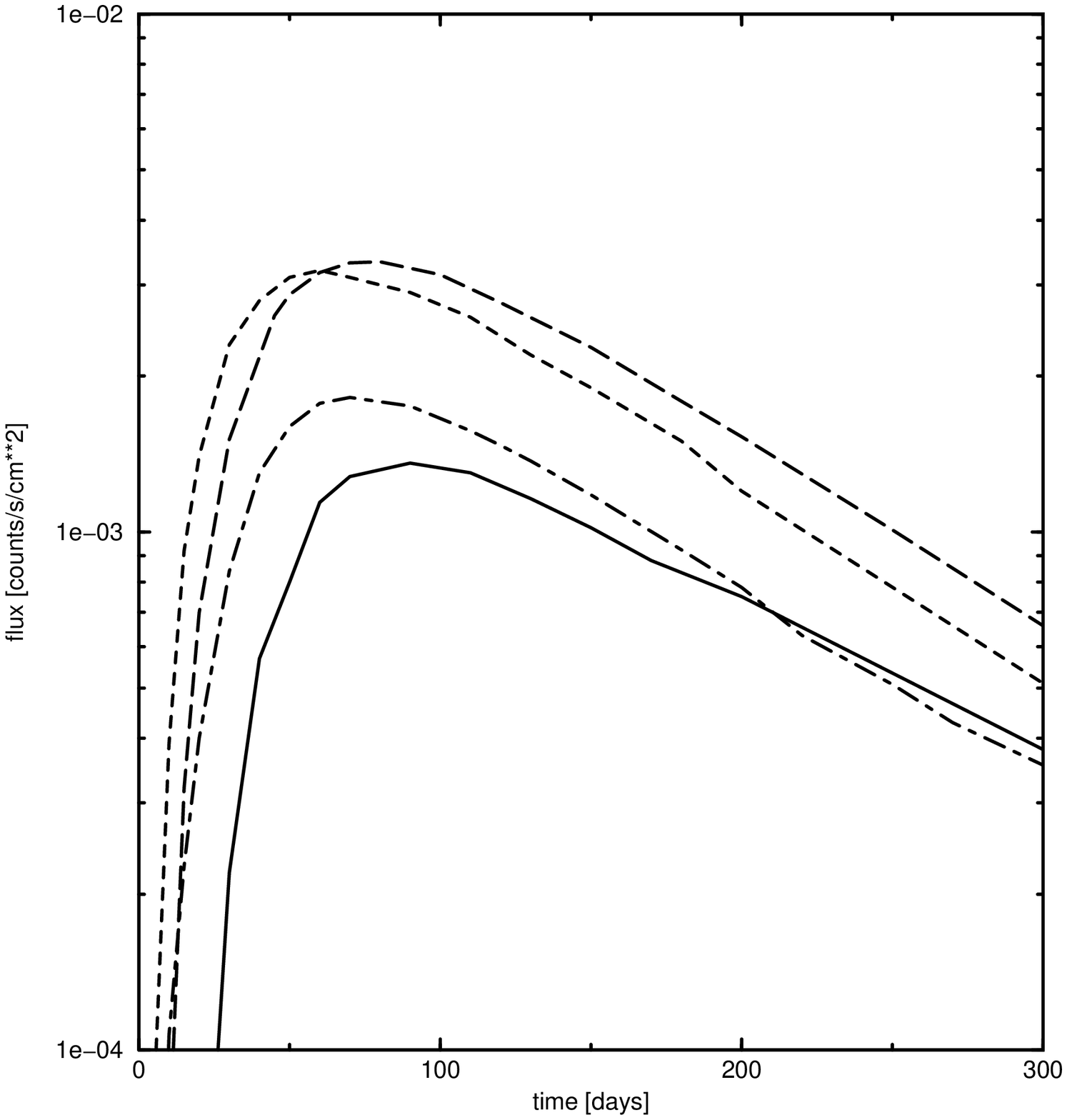,height=8 cm, width=8 cm}
	\end{center}
\caption{Light curves for the  strongest lines assuming a distance of
 1 Mpc. Top panel 
corresponds to the 158 keV line, the central panel to the 847 keV line and 
 the bottom panel to the 1238 keV line.  Solid
 lines correspond to DEF model, long-dashed lines to DEL model, dashed lines to
 DET model and dot-dashed lines to SUB model.}
\label{fig2}
\end{figure}

The evolution of the $\gamma$-ray emission of these models is shown
 by the instantaneous spectra appearing  in Figure \ref{fig1} and by 
 the light curves  of the  strongest
 lines (Figure \ref{fig2}). As expected from the models considered here the
 main properties of these spectra and light curves are compatible with
 those found in the literature for similar scenarios (see results for W7 \cite{Bu90} and DEF 
\cite{Ho94};  WDD2 \cite{Ku95} and N21 \cite{Ho94}; DET1\cite{Ho94}; Model 2 \cite{Wo96}). However,
 not all the properties are comparable since in some of these works either the continuum
 properties or the line profiles are not described.

 Twenty days after the explosion,  the DEF model only shows  a continuum
  component  while the 
 DEL, DET and SUB  already display strong 
lines due to their higher expansion rates.  Lines are particularly intense 
for DET and SUB models since they contain 
  $^{56}$Ni and $^{56}$Co in the outermost shells. The
 efficiency of comptonization to produce continuum at low energies is limited
 in all models by the competing  photo-electric absorption   which 
 imposes a cutoff below 40 -- 100 keV.  The energy of the 
cutoff is determined by the chemical  composition of the external 
layers where most of the emergent continuum is formed at this epoch.
 In DEF and SUB models,   comptonization mainly occurs in regions
   composed by intermediate-mass elements. As a consequence  their continuum
 extends to lower  energies than that of  DET and DEL models since their
 comptonizing layers mostly
 contain  Fe peak elements. Noticeable amounts of $^{56}$Ni are still
 present at this epoch and the   158 keV (Figure \ref{fig1}, top), 480 keV, 750 keV and
 812 keV lines are really strong. The maximum intensities of these lines
 are very model  dependent since at this moment the expansion rate and
the distribution  of   $^{56}$Ni strongly influence the spectrum.
 In models DEL and DET, the high expansion velocities of  $^{56}$Ni and 
 $^{56}$Co produce such  broad lines that in some cases the lines
 emitted by both isotopes blend and  lead to  two peaked light curves
  (Figure \ref{fig2}, center).

Two months after the explosion(Figure \ref{fig1}, center), all the models
 ,except DEF, are 
close to their maximum luminosity (Figure \ref{fig2}). The 
 $^{56}$Ni lines
 have disappeared  while the 122 keV and 136 keV lines produced by $^{57}$Co 
are already 
 visible although faint. At this moment, the line intensities in
 the DEL, DET and SUB models are mainly determined by the total  mass of 
radioactive isotopes,
 while the effect of the expansion rate becomes secondary.  The cutoff energy 
of DEL, DET and SUB  models converge to a value of $\sim$ 70 keV although
 that is still smaller for DEF.

  Four months after the explosion the ejecta are optically thin in all models.
  The continuum is  faint, and  is dominated by a positronium
 annihilation component plus a contribution of photons scattered once.
 This contribution steeply decreases below 170 keV (the energy of a 
backscattered 511 keV) and a step appears at this energy.
 During this phase, the  cutoff is  associated  with  the characteristic
 spectrum of photons  emitted by positronium annihilations which is 
model-independent. Line intensities are now proportional to the mass of the
 respective parent isotopes.

\begin{figure}
	\begin{center}
\epsfig{file=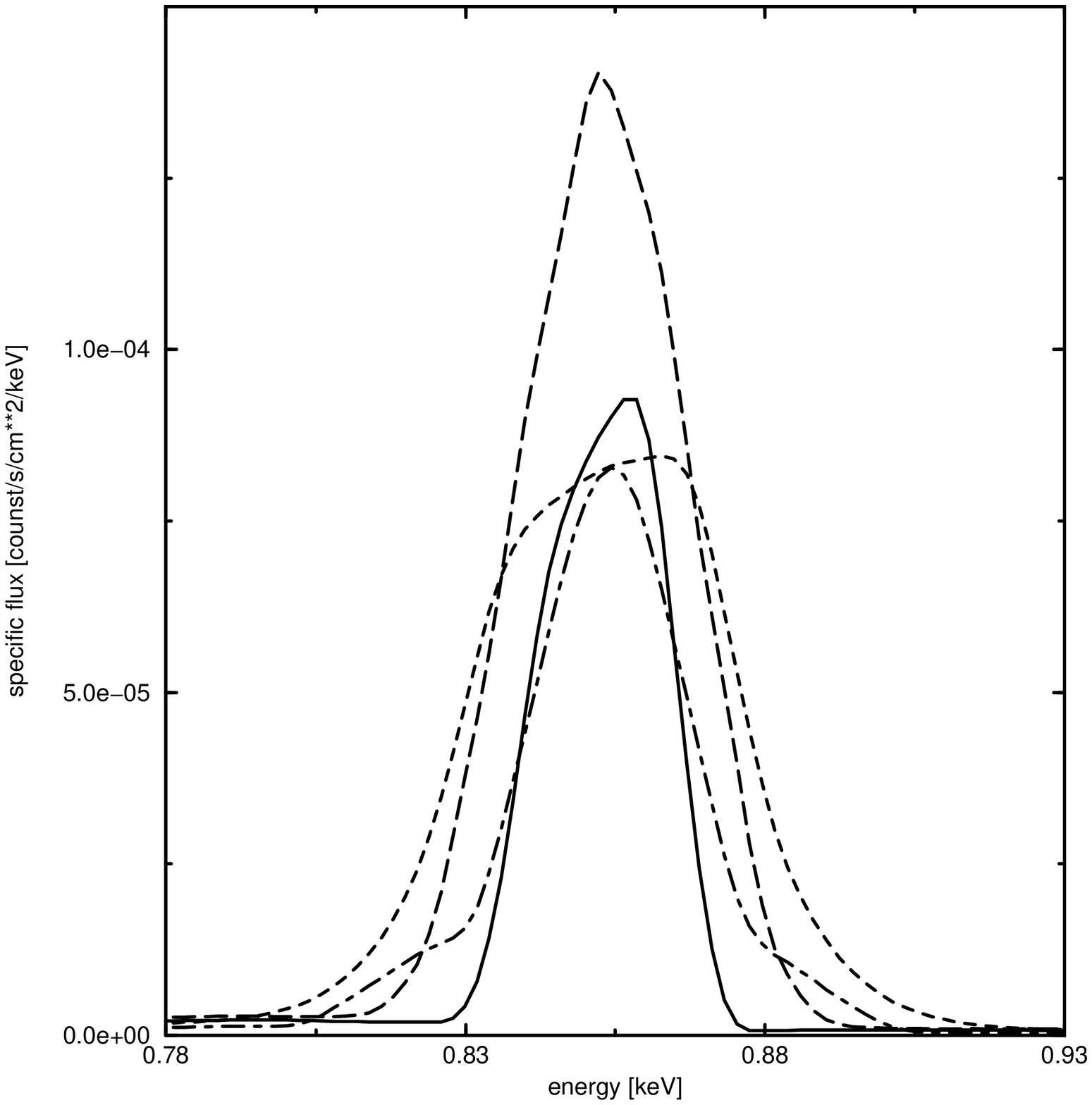,height=8 cm, width=8 cm}
	\end{center}
\caption{The profiles for 847 keV line at 120 days. A distance of  1 Mpc is
 assumed. Lines have the same meaning
as in the previous figures.}
\label{fig3}
\end{figure}

   Line profiles
 reveal  during all their evolution differences  in the velocity distribution 
of their parent isotopes. But  at late epochs, when the models become 
transparent, these profiles give information about all layers of the ejecta (Figure \ref{fig3}).
 The lines corresponding to
DET  are the broadest ones and display a peculiar truncated core produced
 by the absence of $^{56}$Co in the central layers,
 DEL and SUB lines show intermediate widths, while DEF lines are relatively narrow
 due to the low expansion velocity of its envelope. An interesting feature of
 the SUB lines is the presence of line wings which are the result of the typical
 ``sandwich'' distribution of $^{56}$Ni in sub-Chandrasekhar SNeIa (see also,
 Woosley et al., 1996).

\section{Discussion}
In the case of  a close SNIa explosion ($\sim$ 1 Mpc) during the life time of 
INTEGRAL,  high quality $\gamma$-ray spectra could be measured and  detailed 
comparison between observations and theoretical results would be possible. 
However, 
when we consider more
 realistic distances ($>$ 5 Mpc) the information provided by observations
  decreases drastically  and only some outstanding features
  have a chance to be detected (Figure  \ref{fig4}). Our purpose is to
 determine
which are these  spectral features  and  what is the  significance,
 if any, of the observational differences between models.
To perform such study we have taken into account  the theoretical
 spectra computed by us and the current 
estimations of the response of IBIS and SPI instruments included in
  the INTEGRAL mission (Lei 1995; Jean 1995). While SPI provides  extremely 
good spectral capabilities  appropriate for determining line properties,
IBIS  provides a higher effective area that makes it useful
for broad  band measurements of low energy continuum. It is 
important to note that the estimated  sensitivity curves  of these instruments
    do not include the instrumental sources of line
background, which
 could be important at certain energies depending on the materials selected 
for the construction of INTEGRAL. 

We have considered some quantities to summarize the properties of the 
 spectra that are observable at relatively  long distances and  allow to easily determine
 differences between models. These properties are: line intensities,
 continuum intensities and line widths. In all  calculations, an
 integration time of 10$^{6}$ has been assumed.

\subsection{Line intensities}
 Lines emitted by   all the  models are
 broader than SPI instrumental resolution. Therefore,  the intensity of a line
must be measured
 by  integrating   the flux in several instrumental channels forming a
  a  band centered at the line energy. 
This implies on one side,that the  
effective sensitivity of SPI is  reduced as compared
 to the narrow line case, and on the
 other side that it is important to appropriately select the band for the
 integration
 of counts  to increase the significance of observations.
In our calculations we have always adopted the band width which  gives the
 measures 
 with the maximum significance. This band varies not only with the 
 line considered but also with the model.
 As a reference,   in the case of  a gaussian line the maximum significance is 
obtained for a bandwidth of $\sim$ 1.2~FWHM (full width half maximum)  and the
number of counts contained in the band
are  $\sim$ 80 \% of the total line flux. 

In  our models,  the lines with the highest intensities are: 
   511 keV,  847 keV (Figure \ref{fig2}, center), 1238 keV 
(Figure \ref{fig2}, bottom)    812 keV and  158 keV(Figure \ref{fig2}, top).
 Table \ref{Tab2} summarizes the maximum distances at which 
 the 158 keV, 847 keV and 1238 keV lines can be detected. At these 
distances,  the lines would be detected with a
    3~$\sigma$.
Values were computed using the average flux around the time of maximum. 
 In the table also appears  the average number of the
source counts detected in the case of  an explosion at 5 Mpc. In all cases, this
number     falls far below the background level. 
In the table we can also see  that the  similar
 fluxes for DET and DEL lines  correspond to different maximum detection
 distances due to the effect of the line widths in the sensitivity.
In all cases, the  strongest line is 847 keV which is  detectable up to 
distances  between 11 and 16 Mpc.  These distances are noticeably more pesimistic
 than those obtained by Kumagai et al. 1995 (distances from 40 to 56 Mpc) this is probably due to the fact that the broadeness effect has been included   
in our line sensitivity calculations for  SPI. 
The  differences in  the
 fluxes measured at maximum  range a factor of 2 between the different
 models.  DET and DEL models display the highest fluxes, which are almost
identical. In the case of DEF and SUB models, the   fluxes are much lower.
Similar relationships are found for  the fainter 1238 keV line, although in
 this case the fluxes are lower.

In contrast with Burrows \cite{Bu90}, we have preferred to consider the
 158 keV line instead of the 812 keV line (both coming from the decay
 of  $^{56}$Ni) because despite of being fainter its narrower profile makes it
detectable at longer distances for SPI. The behavior of this 
 line is  very model-dependent.
 Despite being fainter than the  847 keV one, 158 keV  is  interesting in order to
 discriminate among  models.
      This line is almost undetectable for the DEF model while   for the SUB and 
DEL models it   shows similar intensities  and it is  detectable up to distances of 
 $\sim$ 7 Mpc. Finally, for a DET model, the 158 keV line 
is even stronger than the 1238 keV one, allowing its detection up to 12 Mpc. 
\begin{table*}
\begin{minipage}{120 mm}
\centering
\caption{Intensities of the strongest lines.}
\label{Tab2}
\begin{tabular}{@{}lccccccccccc}

&\multicolumn{4}{c}{158 keV}&847/1238 keV&\multicolumn{3}{c}{847 keV}&
\multicolumn{3}{c}{1238 keV}\\
MODEL &
t$_{m}$ &
d &
c &
f &
t$_{m}~$  &
d &
c &
f&
d &
c &
f \\
DEF & -- & -- & -- & -- & 95 & 11.2 & 6~10$^3$ &9.2~10$^{-5}$& 7 & 3.5~10$^3$&5.2~10$^{-5}$ \\
DEL  & 25 & 7  & 6~10$^3$&3.1~10$^{-5}$ & 75  & 16.2 & 1.3~10$^4$ &1.9~10$^{-4}$& 12.2 &
 8.4~10$^3$ & 1.3~10$^{-4}$ \\
DET & 14 & 12 & 2~10$^4$&1.0~10$^{-4}$  & 60 & 15.2 & 1.3~10$^4$&1.8~10$^{-4}$ &  10.6 &
 7.7~10$^3$&1.2~10$^{-4}$ \\
SUB & 16 & 8  & 7~10$^3$& 3.8~10$^{-5}$ & 70  & 13.2 & 8.6~10$^3$ &1.2~10$^{-4}$ & 9  & 
4.6~10$^3$ &7.2~10$^{-5}$\\
\end{tabular}

\medskip

 Time of maximum flux {\em (t$_m$)} in days, maximum distance of detection 
{\em (d)} in Mpc  number of counts at 5 Mpc {\em (c)} and flux at 5 Mpc {\em (f)} 
 for the strongest
 lines.

\end{minipage}
\end{table*}

An interesting coefficient is   the relation between  the 847 keV and 
 the 158 keV line fluxes 
$F(847)_{200 days} / F(158)_ {max}$ (see  Table \ref{Tab3}). This value
 provides  information  about the ratio
between the total $^{56}$Ni in the ejecta and the abundance of this isotope in 
the external layers.
 The late emission of 847 keV is a consequence  of the contribution  of all 
$^{56}$Co in the ejecta (produced mostly by the decay of $^{56}$Ni) while
 only the $^{56}$Ni present in the outermost  shells is responsible for the flux in 
the 158 keV line.

\begin{table}
\centering
\caption{$F(847)_{200 days} / F(158)_ {max}$}
\label{Tab3}
\begin{tabular}{@{}lc}
MODEL & 
ratio \\
DEF & 8 \\
DEL  & 2.2 \\
DET & 0.7  \\
SUB & 1.3  \\
\end{tabular}

\medskip

In the coefficients F(847) is taken at 200 days and F(158) at maximum.
 
\end{table}

\subsection{Continuum}
The properties of the continuum can be measured by obtaining the flux in 
fixed bands.
 In this, case the bands are broader than those used 
when measuring lines since the specific flux is lower.
   This means that one cannot  expect to
 measure  properties like  spectral indexes or the  position of the 
cutoff for continuum, but just fluxes in coarse channels.
We have  selected four bands taking into account the  properties
 of the spectra and the detectability requirements.
 The 80--135 keV and  200--540 keV bands  for SPI and the
 40--80 keV and 80--135 keV bands for IBIS. 

 Bands 40 --80 keV and 80 --135 keV  contain information about the
 position of the continuum cutoff of
 the models (which is located in the interval   40--100 keV). At
 these energies  the sensitivity of SPI degrades 
 steeply  preventing its use to detect the continuum  below 80 keV where only
 IBIS is considered.

 The  200--540 keV band contains most of the emission  produced by
positronium annihilation  which is the main source of continuum 
 photons at late times. Due to the fact that  emission of 511 keV photons
occur  only in
 25 \% of  positronium annihilations, 
the significance of the flux in this band is
  higher than that for  the 511 keV line itself (see Table \ref{Tab5}).

 It is important to 
note that, when  measuring  flux  in  broad bands, the 
response of the instrument  vary with the energy.  
To estimate the significance of the  continuum measurements it is necessary to 
convolve the incident flux for each model with the instrumental exposure
 function and to compare the measured counts  with the total background in the 
band.  In this way, significances for the total flux in the band are obtained.
  The  maximum  distances at which the different models
 can be detected and the  average number
 of source counts expected at 5 Mpc are presented in  Table \ref{Tab5}. As for line intensities,
  fluxes have been  averaged over a period of 10$^{6}$ s. The integration 
 time is centered  at 30 days after the explosion, when 
 the continuum is at maximum.
 At low energies the highest significances are obtained for the 80--135 keV
 band, although in this band differences   among the models only range 
$\sim$ 30\%, being  DEF and DET those which have the highest and the 
lowest fluxes respectively.
  Differences between models regarding flux   in the 40 -- 80 keV band are really 
important.  The DEF model  is the only one  that can be detected
 at these energies for  distances $>$ 1 Mpc. The  significances in this band
 for DEF 
are  $\sim$ 20 times higher than those for any other model and make it 
 detectable up to 4 Mpc.

Significances for the 200--540 keV band range a factor of 4 between the model
with lowest fluxes (DEF) and that with the strongest emission (DET). The band
can be detected  up to 11 Mpc for DET  and is  the best candidate for the 
observation of the $\gamma$-ray continuum in SNeIa.
 However, it does not provide with any information about the position
of the cutoff an thus about the elemental composition of the ejecta.

\begin{table*}
\begin{minipage}{120 mm}
\centering
\caption{Properties of  the continuum.}
\label{Tab5}
\begin{tabular}{@{}lcccccccc}

&\multicolumn{4}{c}{IBIS}&\multicolumn{4}{c}{SPI}\\
MODEL &
40--80 d &
40--80 c &
80--135 d &
80--135  c &
80--135 d  &
80--135 c  &
200--540 d  &
200--540 c  \\
 
DEF & 4.2 &  4~10$^4$ & 4 &  5.8~10$^4$ & 3.8 &  7~10$^3$&5 & 6~10$^3$ \\
DEL  & $<$ 0.7 & 10$^3$  & 3.7 &  4.8~10$^4$  &3.5 & 6~10$^4$ & 10&  2.4~10$^4$ \\
DET & $<$0.7  & 10$^3$ & 3.6  &  4.6~10$^4$ & 3.3 & 5.7~10$^4$ & 11 & 2.8~10$^4$ \\
SUB & 1  &  2.5~10$^3$ & 3.9 &   5.3~10$^4$ & 3.6 & 6.5~10$^3$ & 8  & 1.5~10$^4$ \\

\end{tabular}

\medskip

 Maximum distances of detection {\em (d)} and counts at 5 Mpc {\em (c)} for
 different continuum bands. All  bands are in keV and the distances in Mpc.

\end{minipage}
\end{table*}
\subsection{Line widths}
\begin{figure}
	\begin{center}
\epsfig{file=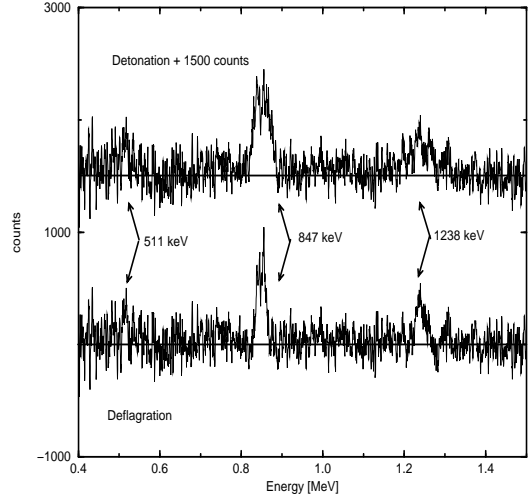,height=8 cm, width=8 cm}
	\end{center}
\caption{Simulated ``observational'' spectra for a detonation and a 
deflagration SNIa at 5 Mpc (integration time= 10$^{6}$s). The detonation
 spectrum is shifted a factor +1500. Obtained by Jean et al. (1995b).}
\label{fig4}
\end{figure}

 In principle, the good spectral resolution of  SPI
should allow to identify many details of the line profiles of SNeIa 
(Figure \ref{fig3})
 and  to  perform model-specific fits to the observations.
 However,  at distances larger  than
$\sim$ 2 Mpc the fluctuations of the background hide the  secondary features
 of the lines, (see Figure \ref{fig4}) and   in the majority of cases
it is only possible to  fit    gaussians. At these distances,
 the width 
 of the lines is the basic property of their profiles that we can measure.
 Although for  DEF, DEL, DET and SUB  models  lines  are not exactly
 gaussians, the  difference between the FWHM of 
 their  theoretical  profiles  and  that of the corresponding  gaussian fit
is,  in all cases,   below $\sim$ 3 \%. These errors are negligible 
compared with 
  observational uncertainties and hence gaussian fitting is a good  technique to
 measure observational line widths. In what follows, we take  the width of a line as the
 FWHM of its gaussian fit.

 Our purpose is to estimate the maximum distances at which it is possible
 to  obtain   significant width measurements for the 
 $\gamma$-ray lines predicted in our simulations.  To do this, it is necessary
 to determine the errors associated to the observational  fits.
In opposition to what happens with   measures of line
 intensities and continuum fluxes, the   errors expected  for widths 
cannot be analytically inferred from the instrumental properties
 since the average deviations of width  measurements  have complex dependencies on the
 properties  of background fluctuations.
To cricumvent this difficulty   we have  applied a  bootstrap method 
  by  simulating an statistically significant number  of ``observations'' of 
 the lines emitted by  the different models, placed at  several explosion 
distances, in order to directly determine the distribution of the  errors 
which appear when fitting their widths.

The procedure we apply is the following one.
 Taking  a line and a explosion distance, the  theoretical line profile at that
  distance is  convolved with the exposure function of SPI (this function
 expresses the efficiency of the instrument to detect source photons) and  with
 a gaussian  that accounts for  the limited resolution of SPI. 
The result of these two calculations is  the signal that would be
 generated by the source line on the instrument  in total counts per  channel. 
 However,   in  actual observations the source  signal
is additionally  contaminated by the 
 fluctuations due to the background subtraction
 procedure whose influence can only be 
predicted in a statistic way.  To evaluate such effect,  we have  generated
for each convolved line a 
 set of 
hypothetical observations where   random background fluctuations have been added 
to  the source signal.
 These fluctuations  are generated with  a Monte Carlo 
 procedure   that takes into account  the background  properties   predicted
 for SPI.  A gaussian has been fitted  to
 every "observation"   by using the   $\chi^{2}$ technique as it would
 be  actually done in a real observation and finally 
 a line width has been  obtained.
In each set    more than 300 observations have been simulated.

\begin{figure}
	\begin{center}
\epsfig{file=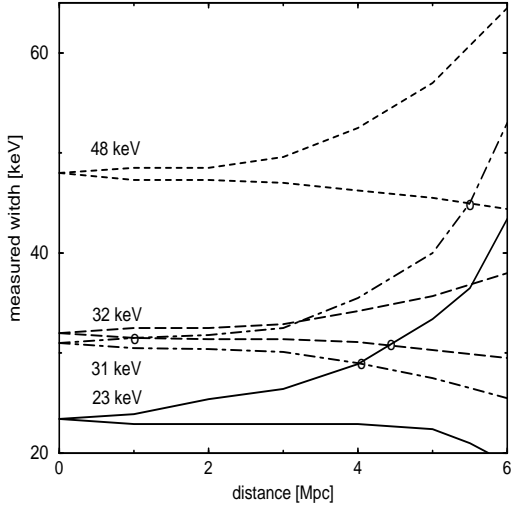,height=8 cm, width=8 cm}
	\end{center}
\caption{ Pairs of curves  containing 90 \% of  observational line widths 
  as a function of distance for the 847 keV
 line  (120 days after the explosion). The dispersion of the measures is due to the
 fluctuations of the background. Solid line corresponds to DEF model, 
dotted line to DEL model,
dashed line to DET model and dot-dashed line to SUB model. Also indicated appears
the original line width for each model.}
\label{fig5}
\end{figure}

With this procedure  it is possible to  determine the 
 distribution of errors  associated to the measurement of the  width for any line, 
for any model and explosion distance. Calculations have only  been performed  for
 the 847 keV line, since it  has the maximum chance of being detected. The 
observations were taken  120 days after the explosion.
The results obtained  are summarized in Figure \ref{fig5}. 
 where for each model a pair of curves is displayed. At
 every  explosion distance  the  pair  defines an interval of possible measured 
widths which  contains the values
 that would be obtained by   90 \% of  observers measuring the same
line at the same distance (90 \% dispersion bar). 
In the figure  it can be  appreciated that  the  
 dispersion of the  measures is 0 for an explosion at distance 0 but it  steeply 
grows with the explosion distance, being 
 larger for lines with low fluxes.
This is  particularly important for the 
  SUB  and DEF models, since they  have   the lowest luminosities.
For all models  the distribution of hypothetical measures
 is skewed. That is,
 the observations are  not symmetrically spread around the original line width
but there is a tendency to  measure   widths larger than the original values which
 are indicated in the figure. 
 As the possible  errors become more important the significance of a measure
 decreases. Hence, it is necessary to  adopt a  quantitative criterium 
 which establishes the maximum distance 
 at which a measurement of a line width has physical  meaning.
We take this distance at the point at which the width of the
dispersion bar for a line  equals its original width.
Assuming this definition the     distances  are:
$\sim$ 5.5 Mpc, $\sim$ 8 Mpc
 $\sim$ 7.5 Mpc and $\sim$ 6 Mpc for DEF, DEL, DET and SUB respectively.

But  more important than  determining  when the width of a line can be
 measured, is to know when this measure will  be  useful to discriminate among
 the different models.
 Figure \ref{fig5} provides   the information
necessary to  reject or identify models from observations using measured  widths. Taking 
the measured width and the explosion distance for a given observation an observer
 can  reject all the models whose associated  pair of curves in the figure
  do not contain the measure. By doing this we  ensure
  a probability $>$ 90 \% of correct model  rejection.
 In the same way, if the pair of curves corresponding to  only one of the  models
 contains the given measure,   the observer can identify this model as observed.
 This  would lead to a correct model identification 
in more than   90 \% of cases.
 However, the confidence of the method decreases when we consider distances
where the 90 \% dispersion bars of two or more models   overlap.
  For any pair of models, the point of intersection  between the 
respective pairs of  curves places the distance  below  of which it can be
 assured that 
 model discrimination with the use of witdh measurements will be correctly 
performed in more than 90 \% of observations. We take this distance as
 the maximum distance reasonable  for  model discrimination by means of
 width measurements.
  In Table \ref{Tab6} we summarize
 these maximum distances for the different combinations of models.
Line widths for SUB and DEL models are  indistinguishable for distances larger
 than 1 Mpc. However, in the
 remaining cases the differences between each model pair are noticeable and 
it is possible to discriminate among them up to distances of the order of
4 to 7 Mpc. Particularly interesting are the
differences between DEL and DET models which otherwise have very similar line 
intensities (Table \ref{Tab2}).

\begin{table}
\centering
\caption{Line width measurements.}
\label{Tab6}

\begin{tabular}{@{}lcccc}
MOD/MOD &
DEF &
DEL &
DET &
SUB \\
DEF & - & 4.5 Mpc  & 6 Mpc & 4 Mpc \\
DEL  & 4.5 Mpc & -  & 7 Mpc & $<$ 1 Mpc  \\
DET & 6 Mpc  & 7 Mpc & - &  5 Mpc \\
SUB & 4 Mpc  &  $<$ 1 Mpc & 5 Mpc \\
\end{tabular}

\medskip

 Distances at which the line widths allow to discriminate  different
 models with at least 90 \% probability. 

\end{table}

 If we had  adopted a level of confidence lower than 90 \% it would have been
possible to  define longer distances for model discrimination. However,
due to the steep increase of the dispersion at the distances the limits obatined
would not be appreciably longer.

\section{Nuclear excitation}
  Radioactivity  is  not the only source of
 $\gamma$-ray emission in  SNIa explosions.   The interaction of the high
 velocity ejecta with the surroundings    can induce  emission of 
$\gamma$-rays due to    nuclear excitations and 
spallation reactions. For
  comparison purposes  with the radioactive decay mechanism,  we have
  investigated the importance of such  emission source in the case of
 SNeIa.   Two situations have been considered:  the interaction of the
 high velocity ejecta with the stellar wind
emitted by the supernova progenitor (case of the explosion of a symbiotic 
 binary, for instance) and the interaction  with a relatively dense 
interstellar medium.

In  the  symbiotic binary scenario  we have  assumed a
  system  formed by a red giant, a white dwarf 
 and an  associated  stationary wind.  We have adopted a  total mass of 
   3 \Msol, an orbital period of  1 yr and  a separation of 1.7 AU.
 The composition of the wind is taken to be  solar   while that
 of the ejecta is taken from the DEF, DEL, DET and SUB models respectively.
 If we assume an accretion efficiency of  10 \%,   an accretion rate
 appropriate for a sub-Chandrasekhar supernova ($\sim$ 10$^{-8}$ \Msol 
yr$^{-1}$ ) 
 and a wind velocity of 10 km/s, the density profile of the wind is:
$$
\rho~~=~~\frac{10~ \dot{M}_{accr}}{4 \pi r^{2} v_{wind} }~~=~~\frac{2.5~10^{11}}
{r^2}~
gr/cm^{3}$$
where r is the distance to the center of the system. In this scenario we have
 not considered any energy losses of the ejecta 
since the $\gamma$-ray emission
 almost disappears after 10--15 yrs, much earlier than the beginning of the
 Sedov phase of the expansion.

In the  case of an  explosion  in   a  ISM region,  
the  density is  a free parameter. 
 The average density of the ISM in the galactic plane is $\sim$ 1 cm$^{-3}$ but 
 regions with  densities  ranging 10 -- 100 cm$^{-3}$ are not rare.
 In principle, SNeIa
 are not specifically associated with dense ISM regions but  they could
 meet such regions  when crossing the galactic plane.
 In this scenario, the ISM region is considered to be unbounded and hence the 
emission will be  persistent.  In this case,
 the effects of braking  have to be taken in to account since they are
 the main  responsible  for
  the late decay of the light curve.  They are
  caused by ionization and excitation of the ISM atoms and can be described
 by the typical Bethe-Block equation which takes into account the composition of
 the ejecta and of the ISM.  In this scenario both,
the intensity of the emission and the  stopping power  are proportional
 to the density of the medium   and lead to the following
scaling law for the light curves as a function of the ISM density:
$$L \left ( t,n \right )~=~\frac{n}{n_0}~L \left( t~\frac{n}{n_0},n_0 \right)$$
where n and n$_{0}$ are different ISM densities and t is the time.

We have computed the main light curves  induced by the  DEL, DET and SUB 
models
 in both
 scenarios.
    The DEF model
does not eject  mass  with  energies above the threshold of the
  excitation reactions
 and it can be ignored.  The method used to  compute the light curves is described in 
Fields et al. (1996). The basic equation expressing  the intensity of
a line (l) generated by the interaction  of species (i) in the ejecta with  the
 species (j) in the surrounding medium (with homogeneous composition) is:
$$
L_{l}~=~\frac{1}{4 \pi d^2 A_{i} m^2_p} \int_{0}^{E_{max}}~n_{j}(E,t) 
~\sigma^{l}_{ij}(E)~
X_{i}(E)~\frac{dM}{dE}~v(E) dE
$$
where the specific kinetic energy across the ejecta is taken as the
 integration variable, d is the distance to the observer $X_{i}$ is the mass
 fraction of 
species i, $n_{j}$ is the number density of the (j) species in the medium,
  v(E) is the velocity of a layer of the ejecta with specific kinetic
energy E and $\sigma^l_{ij}(E)$ is the energy-dependent cross section of the
 excitation reaction. Contrary to what was done in Fields et al.,
 in this work braking effects are included and thus the profile of the
 kinetic energy of the ejecta as a function of the mass coordinate evolves
 with time.
  Cross sections were taken from Ramaty et al.(1979) and
 include  direct photon emission and de-excitation of spallation products.

\begin{figure}
	\begin{center}
\epsfig{file=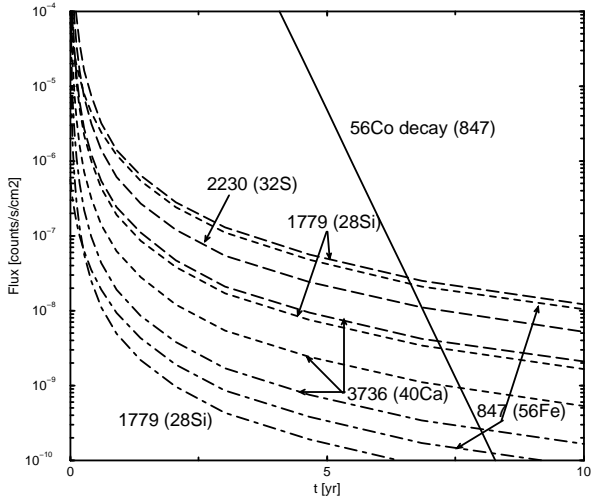,height=8 cm}

	\end{center}
\caption{Evolution of the flux  of the  main excitation lines for a SNIa
 explosion in a symbiotic binary, compared with 847 keV line produced by 
the decay of $^{56}$Co (10 kpc). The lines have the same meaning as in  
 previous figures.}
\label{fig6}
\end{figure}

\begin{figure}
	\begin{center}
\epsfig{file=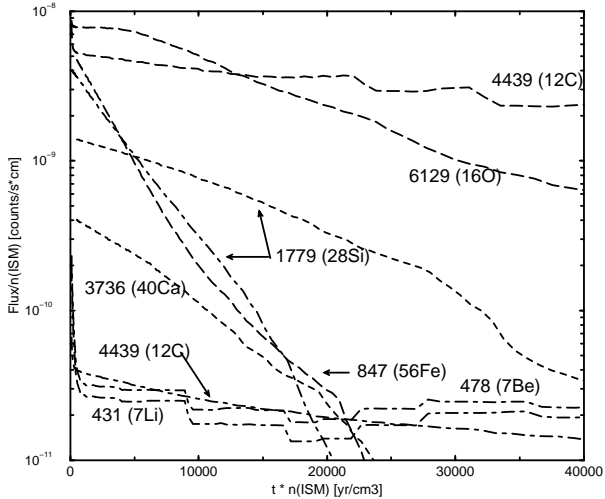,height=8 cm}

	\end{center}
\caption{Evolution of the flux the main excitation lines produced by the explosion of
a SNIa in a dense ISM (10 kpc). X and Y axes can be rescaled as a function of
the ISM density. Lines have the same meaning as in previous figures.}
\label{fig7}
\end{figure}

The results of these calculations are summarized in Figures 
\ref{fig6} and  \ref{fig7} (this figure is valid for all ISM densities due to
 the scaling law previously explained) where the main light curves are displayed.
 The isotopes responsible for the strongest lines are: DEL ( $^{56}$Fe, $^{28}$Si, 
$^{40}$Ca), DET ( $^{56}$Fe, $^{28}$Si, $^{40}$Ca, $^{32}$S, $^{12}$C, 
$^{16}$O) and SUB  ( $^{56}$Fe, $^{28}$Si, $^{40}$Ca ). The main conclusion is that
 the excitation mechanism is in general negligible.
 The emission is  extremely weak, even when   galactic supernovae are considered,
 and in all cases, the luminosity coming from the radioactive decay 
 would overwhelm the excitation lines during  at least
  the first ten years (Figure \ref{fig6}). In both scenarios  the
 models  with the strongest and the faintest emission are the DEL and
 SUB ones respectively.


In the symbiotic scenario, the strongest line (4439 keV), can be detected  at
 a maximum distance of 1 kpc 10 years after the explosion and just 100 pc 
after 100 years.
This means, that
  SNIa explosions in symbiotic systems can be ruled out as a
    source of detectable  emission produced by nuclear excitation.

In  the ISM scenario, the emission is also extremely low although
  the persistence of  lines increases somewhat the chances of
 detection. If a ISM density of 
100 cm$^{-3}$ and the DEL model are considered, the 4439 keV line shows a nearly constant
 intensity  during 150 years before starting to  decay. During this
 period, the line
 could be detected by INTEGRAL for explosions closer than $\sim$ 1.5 kpc. If we assume 
an ISM density of 10 cm$^{-3}$ cm, the line emission lasts for 1500 yrs but
 in this  case it is only detectable at distances $<$ 500 pc. Thus, 
 nuclear excitation can be safely neglected in all cases in comparison with
 radioactive decay as a source of $\gamma$-rays in SNIa explosions. Only
 in the improbably case that one of the recent  close SNeIa (SN 1006, SN Tycho)
 remmants was associated with a dense
 ISM region, its excitation emission could be detected by INTEGRAL.

\subsection{Conclusions}
The simulations performed here show  clear 
differences  between  $\gamma$-ray spectra of deflagration, detonation, 
delayed detonation and sub-Chandrasekhar models of  SNeIa. 
These  differences are particularly
important at early times, when the total amount of
radioactive 
isotopes and  their distribution,   expansion rate and  
composition of the ejecta affect the
$\gamma$-ray emission. After the maximum, differences  are only related with
 total mass of $^{56}$Co and its velocityñ distribution. The main
 differences between the spectra
 are found in the intensity and profiles of the
 strongest lines  coming both from $^{56}$Ni and  $^{56}$Co decays, as 
well as in
 the intensity, shape  and cutoff of the low energy continuum.

Since the    level of background is high  and  the instrumental  capabilities
 are limited, 
 not all the differences present in the emergent spectra of the models 
are actually  observable by an instrument like INTEGRAL. 
Due to the  intrinsic width of SNIa lines, the
 effective sensitivity of SPI to detect  them is reduced by a factor 
$\sim$ 3 -- 4 respect to the narrow line case. The strongest SNIa spectral feature 
is the 847 keV line 
 which can be measured with  3 $\sigma$ for distances up to 11 to 16 Mpc. 
Important 
 variations appear for the fainter 158 keV line
 which allow to discriminate 
 models up to 6 Mpc. This line is  detectable for the a detonation SNIa
 up to 11 Mpc.
The continuum must  be measured  using broad bands. Emission in the 80 -- 135 keV
 could be  observed (3 $\sigma$) at 4 Mpc although in this case there are only
 small differences 
among the different models. The flux in the  40 -- 80 keV band is a peculiar
 feature of deflagration SNIa and
 could be observed with IBIS  up to  4 Mpc. Concerning line profiles,
 in principle they can 
 be 
 resolved by SPI. However the limited sensitivity reduces the information that can
 be  obtained about the shapes and only line widths can be measured at reasonable distances.
We have shown here that   it is possible to 
discriminate between any pair of models for distances up to 4 -- 6 Mpc by
measuring  the width of 847 keV line except
 in the cases  DEL and SUB.

From these results, the observation of a SNIa explosion is possible with 
INTEGRAL for any event closer 
than 11 Mpc. If the explosion mechanism of SNIa is a detonation they could be observed at
distances as long as 16 Mpc.
However, only if a SNIa explodes at a distance of
 $\sim$ 6 Mpc or less, INTEGRAL could obtain   enough 
information  to univocally discriminate between the $\gamma$-ray
  emission patterns corresponding to DEF, DEL, DET and SUB models. In fact,
only  a single recent SNIa (SN1986G)  fulfilled this requirement.

Finally, it has been shown that the nuclear excitation mechanism can be 
ruled out 
as a source of detectable $\gamma$-ray emission. In any of the proposed
scenarios this emission is totally dominated by the radioactive decay emission
 for more than ten years. After that,  in the symbiotic scenario the luminosity 
decays very quickly. Only a SNIa event  that had happened  in the solar
 neighborhood during the last few centuries and that was associated  with 
a high density  ISM could be detected as a fossil $\gamma$-ray emitter.

\section*{ACKNOWLEDGMENTS}

We wish to  thank Eduard Bravo and Jordi Jos\'e, who  provided us with 
 initial models for our calculations. Peter von Ballmoos  discussed with 
us fundamental questions about the instrumental response of SPI  and
gave us an ``observational''  perspective of the problem. 
 This work has been financed by the projects: 
CICYT  (ESP95-0091) and CIRIT (GRQ-8001).

\label{lastpage}
\end{document}